\documentclass[conference]{IEEEtran}
\usepackage{amsmath}

\usepackage{xcolor} 
\usepackage{verbatim}
\usepackage{amsfonts} 
\usepackage{stmaryrd} 
\usepackage{nicefrac} 
\usepackage[normalem]{ulem} 
\usepackage{tipa}
\usepackage{extraipa} %
\usepackage{enumerate}
\usepackage{hyperref}

\newcommand\FN[1] {{\color{red}\footnote{\color{red}{#1}}}}

\newcommand\Section[2][\empty]
	{\ifx\empty#1\section{#2}
	\else\section{#2}\label{#1}\marginpar{\tiny#1}\fi}
\newcommand\SSection[2][\empty]
	{\ifx\empty#1\subsection{#2}
	\else\subsection{#2}\label{#1}\marginpar{\tiny#1}\fi}
\newcommand\SSSection[2][\empty]
	{\ifx\empty#1\subsubsection{#2}
	\else\subsubsection{#2}\label{#1}\marginpar{\tiny#1}\fi}
\newcommand\BE[1] {\begin{equation}\label{#1}\HangLeft[\qquad]{\tiny #1}~}
\newcommand\EE {\end{equation}}

\renewcommand\FN[1] {\relax}
\renewcommand\Section[2][\empty]
	{\section{#2}\ifx\empty#1\else\label{#1}\fi}
\renewcommand\SSection[2][\empty]
	{\subsection{#2}\ifx\empty#1\else\label{#1}\fi}
\renewcommand\SSSection[2][\empty]
	{\subsubsection{#2}\ifx\empty#1\else\label{#1}\fi}
\renewcommand\BE[1] {\begin{equation}\label{#1}}
\renewcommand\EE {\end{equation}}

\newcommand\Thm[1] {Thm.~\ref{#1}}
\newcommand\Sec[1] {\S\ref{#1}}

\newcommand\TN{\ensuremath{T\kern-.1em,\kern-.1emN}}

\newcommand\UI {\ensuremath{{\cal U}}}
\newcommand\UIK {\ensuremath{{\kern.1em\cal U}}}
\newcommand\UIT {\ensuremath{{\cal U}_{\kern.07emT}}}
\newcommand\PDF {\textsc{pdf}}
\newcommand\Min {\textsf{\small min}}
\newcommand\Inf {\textsf{\small inf}}

\newcommand\Bag {\ensuremath{\mathbb B}}
\newcommand\DB {\Bag\kern-.03emR}
\newcommand\Real {{\mathbb R}}
\newcommand\RealNN {{\mathbb R}^\geq}
\newcommand\Dist {{\mathbb D}}
\newcommand\Meas {{\mathbb M}}

\newcommand\Fun {\mathbin{\rightarrow}}
\newcommand\Size {\#}
\newcommand\SD {\mathbin{\bigtriangleup}}
\newcommand\In {{:}\,}

\newcommand\XX {\ensuremath{{\cal X}}}
\newcommand\YY {\ensuremath{{\cal Y}}}
\newcommand\WW {\ensuremath{{\cal W}}}

\newcommand\GeoD {\ensuremath{G}}
\newcommand\Geo {\ensuremath{G}}
\newcommand\Q[1] {\HangRight{\quad#1}}
\newcommand\HangLeft[2][\empty]
	{\ifx#1\empty\makebox[0pt][r]{#2}\else\makebox[0pt][r]{#2#1}\fi}
\newcommand\HangRight[2][\empty]
	{\ifx#1\empty\makebox[0pt][l]{#2}\else\makebox[0pt][l]{#1#2}\fi}
\newcommand\Eqn[1] {(\ref{#1})}
\newcommand\Itm[1] {(\ref{#1})}
\newcommand\Wide[1] {~~~#1~~~}
\newcommand\Defs {:=}
\newcommand\WideDefs {\Defs\quad}
\newcommand\Max {\textsf{max}}
\newcommand\NF[2] {\nicefrac{#1}{#2}}
\newcommand\WRT {wrt.}

\newcommand\UNIF {\raisebox{.3ex}{\footnotesize\ensuremath{\Unif}}}
\newcommand\Unif {\odot}
\newcommand\Times {{\times}}
\newcommand\MFun {\mathbin{\rightarrowtriangle}}
\newcommand\JD[2] {#1\kern.1em{\triangleright}\kern.1em#2}
\newcommand\Hyp[1] {[#1]}
\newcommand\JDH[2] {\Hyp{#1\kern.1em{\triangleright}\kern.1em#2}}

\newcommand\Exp[2] {{\cal E}_{#1}\,{#2}}
\newcommand\Loss[3]{\textsf{\small\$}(#1,#2,#3)}
\newcommand\NStep {$N$\kern-.1em-step}
\newcommand\NStepped {$N$\kern-.1em-stepped}

\newcommand\Dd {\mbox{\sc\small d}}
\newcommand\DD {\ensuremath{\,{\textsf{\small d}_D}}}
\newcommand\DX {\kern.1em{\rm d}}
\newcommand\DM {\mathbin{\cdot}} 
\newcommand\Vep {\varepsilon}
\newcommand\VEP {\ensuremath{\Vep}}
\newcommand\DP {\textit{DP}}
\newcommand\VDP {\VEP-\DP}
\newcommand\LV {\ensuremath{{{\textit{L}^\varepsilon}}}}
\newcommand\LTV {\ensuremath{\textit{\L}^\varepsilon}}
\newcommand\LTVN[1][\empty]{\ensuremath{{\spreadlips{L}}^\varepsilon_{\ifx#1\empty N\else#1\fi}}}

\newcommand\Lap {\ensuremath{L}}

\newcommand\Dirac[1] {\textrm{\boldmath$\delta$}_{#1}}

\newtheorem{theorem}{Theorem}
\newtheorem{lemma}[theorem]{Lemma}
\newtheorem{corollary}[theorem]{Corollary}
\newtheorem{definition}[theorem]{Definition}

\newcommand\Lem[1] {Lem.~\ref{#1}}
\newcommand\Def[1] {Def.~\ref{#1}}

\newcommand\Fig[1] {Fig.~\ref{#1}}

\newcommand\Floor[1]{\lfloor #1 \rfloor}
\newcommand\NFloor[2]{\lfloor #1 \rfloor_{#2}}

\newcommand\Apply {\mathbin{\raisebox{.1em}{\kern.05em\scriptsize$\rangle$}}}
\newcommand\ApplyR {\mathbin{\raisebox{.1em}{\kern.05em\scriptsize$\langle$}}}

\renewcommand\Ref {\mathrel\sqsubseteq}

\newcommand\Kant{{\mathbb W}}

\DeclareMathAlphabet{\mathpzc}{OT1}{pzc}{m}{it}
\newcommand{\len}{\mathpzc{len}}

\usepackage{graphicx}
\usepackage{subcaption}  

\newenvironment{Reason}{\vspace{-.0em}\begin{tabbing}\hspace{2em}\= \hspace{1cm} \= \kill}
    {\end{tabbing}\vspace{-1em}}
\newcommand\Step[2] {#1 \> $\begin{array}[t]{@{}llll}#2\end{array}$ \\}
\newcommand\StepR[3] {#1 \> $\begin{array}[t]{@{}llll}#3\end{array}$
    \` {\RF \makebox[0pt][r]{\begin{tabular}[t]{r}``#2''\end{tabular}}} \\}
\newcommand\WideStepR[3] {#1 \>
    $\begin{array}[t]{@{}ll}~\\#3\end{array}$ \`
    {\RF \makebox[0pt][r]{\begin{tabular}[t]{r}``#2''\end{tabular}}} \\}
\newcommand\RF {\small}

\begin{document}
\title{
The Laplace Mechanism has optimal utility \\ for differential privacy over continuous queries
}
\author{\IEEEauthorblockN{Natasha Fernandes\IEEEauthorrefmark{1}\IEEEauthorrefmark{3}\IEEEauthorrefmark{4},
Annabelle McIver\IEEEauthorrefmark{1},
Carroll Morgan\IEEEauthorrefmark{2}\IEEEauthorrefmark{3}}
\IEEEauthorblockA{\IEEEauthorrefmark{1}Department of Computing,
Macquarie University, Sydney}
\IEEEauthorblockA{\IEEEauthorrefmark{2}School of Computer 
        Science and Engineering,
        UNSW, Sydney}
\IEEEauthorblockA{\IEEEauthorrefmark{3}Data61, CSIRO, Sydney}
\IEEEauthorblockA{\IEEEauthorrefmark{4}Inria and \'{E}cole Polytechnique, IPP, France}
\thanks{\copyright 2021 IEEE. Personal use of this material is permitted. Permission from IEEE must be obtained for all other uses, in any current or future media, including reprinting/republishing this material for advertising or promotional purposes, creating new collective works, for resale or redistribution to servers or lists, or reuse of any copyrighted component of this work in other works.}
}

\IEEEoverridecommandlockouts

\maketitle

\begin{abstract}
Differential Privacy protects individuals' data when statistical queries are published from aggregated databases: applying ``obfuscating'' mechanisms to the query results makes the released information less specific but, unavoidably, also decreases its utility. Yet it has been shown that for \emph{discrete} data (e.g.\ counting queries), a mandated degree of privacy and a reasonable interpretation of loss of utility, the Geometric obfuscating mechanism is optimal: it loses as little utility as possible [Ghosh et al.\cite{Ghosh:12}]. 

For \emph{continuous} query results however (e.g.\ real numbers) the optimality result does not hold. Our contribution here is to show that optimality is regained by using the \emph{Laplace} mechanism for the obfuscation. 

The technical apparatus involved includes the earlier discrete result [Ghosh op.\ cit.], recent work on abstract channels and their geometric representation as hyper-distributions [Alvim et al.\cite{Alvim:2014aa}], and the dual interpretations of distance between distributions provided by the Kantorovich-Rubinstein Theorem.
\end{abstract}


\Section[s1349]{Introduction}
\SSection[s1517]{The existing optimality result, and our extension}

Differential Privacy (\DP) concerns databases from which (database-) queries produce statistics: a database of information about people can be queried e.g.\ to reveal their average height, or how many of them are men.
But a risk is that from a \emph{general} statistic, specific information might be revealed about individuals' data: whether a specific person is a man, or his height, or even both.
Differentially-private ``obfuscating'' mechanisms diminish that risk by perturbing their inputs (the raw query results) to produce outputs (the query reported) that are slightly wrong in a probabilistically unpredictable way. That diminishes the personal privacy risk (good) but also diminishes the statistics' utility (bad).

The existing optimality result is that for a mandated differential privacy parameter, some $\Vep{>}0$, and under conditions we will explain, the \emph{Geometric} obfuscating mechanism $\Geo^\Vep$ (depending on \VEP) loses the least utility of \emph{any} \VEP-Differentially Private oblivious obfuscating mechanism for the same \VEP, that loss being caused by her having to use the perturbed statistic instead of the real one \cite{Ghosh:12}.

A conspicuous feature of \VDP\ (that is \VEP-differential privacy) is that it is achieved \emph{without} having to know the nature of the individual's privacy that it is protecting: it is simply made ``\VEP-difficult'' to determine whether \emph{any} of his data is in the database at all. Similarly the minimisation of an observer's loss (of utility) is achieved by the optimal obfuscation \emph{without} knowing precisely how the obfuscation affects her: instead, the existence of a ``loss function'' is postulated that monetises her loss (think ``dollars'') based on the raw query (which she does not know) and the obfuscated query (which she does know) --- and optimality of $\Geo^\Vep$ holds wrt.\ \emph{all} loss functions (within certain realistic constraints) and \emph{all} (other) \VDP\ mechanisms $M^\Vep$.

\textsc{in summary}: The existing result states that the \VDP\ Geometric obfuscating mechanism $\Geo^\Vep$ minimises loss of utility to an observer when the query results are discrete, e.g.\ counting queries in some $(0..N)$, and certain reasonable constraints apply to the monetisation of loss. But the result does not hold when the query results are continuous, e.g.\ in the unit interval $[0,1]$. \textbf{We show that optimality is regained by using the \VDP\ \emph{Laplace} mechanism $\Lap^\Vep$.}

\Section[s1225]{Differential privacy, loss of utility\\ and optimality}
\SSection[s125718]{Differential privacy}
Differential privacy begins with a database that is a multiset of \emph{rows} drawn from some set $R$ \cite{DworkMNS06}; thus the type of a database is $\DB$ (using ``\Bag'' for ``bag''). A query $q$ is a function from database to a query-result in some set \XX, the input of the mechanism, and is thus of type $\DB{\Fun}\XX$.

A distance function between databases $\Dd\In\DB{\times}\DB\Fun\Real$ measures how different two databases are from each other.
Often used is the \emph{Hamming} distance $\Dd_H$,\,%
\footnote{The Hamming distance is also known as the \emph{symmetric} distance.}
which gives (as an integer) how many whole rows would have to be removed or inserted to transform one database into another: given two databases $b_1,b_2\In\DB$ we define $\Dd_H(b_1,b_2)$ to be the size $\Size(b_1\SD b_2)$ of their (multiset-) symmetric difference.
Thus in particular two databases that differ only because a row has been removed from one of them have Hamming distance 1, and we say that such databases are \emph{adjacent}. 

We define also a distance function (metric) between \mbox{--for} the moment-- discrete distributions $\Dist\YY$ over a set of observations $\YY$ the output of the mechanism.
Given two distributions $\delta_1,\delta_2$ on $\YY$, their distance $\DD(\delta_1,\delta_2)$ (for ``Dwork'') is based on the largest ratio over all $Y{\subseteq}\YY$ between probabilities assigned to $Y$ by $\delta_1$ and $\delta_2$ --- it is
\BE{e0933}
    \DD(\delta_1,\delta_2) \WideDefs \Max_{\,Y{\subseteq}\YY}~|\ln(\delta_1(Y)/\delta_2(Y))| 
\EE
where $\delta(Y)$ is the probability $\delta$ assigns to the whole subset $Y$ of $\YY$, and the logarithm is introduced to make the distance satisfy the triangle inequality that metrics require.\,%
\footnote{This distance is also known as ``max divergence''.}

Following the presentation of Chatzikokolakis et al.\ \cite{Chatzi2013}, once we have chosen a metric \Dd\ on databases, we say that a mechanism $M$ achieves \VEP-Differential Privacy \WRT\ that \Dd\ and some query $q$, i.e.\ is \VDP\ for $\Dd,q$, just when
\BE{e0944}
    \begin{array}{l}
        \mbox{for all databases $b_1,b_2$ in $\DB$ we have} \\
        \DD(M(q(b_1)),M(q(b_2))) \Wide{\leq} \Vep\DM \Dd(b_1,b_2) \Q.
    \end{array}    
\EE
In the special case when \Dd\ is the Hamming distance $\Dd_H$, the above definition becomes
\BE{e1057}
    \begin{array}{l}
        \mbox{for all databases $b_1,b_2$ in $\DB$ with $\Dd_H(b_1,b_2){\leq}1$}, \\
        \mbox{i.e.\ that differ only in the presence/absence of a single row}, \\
        \mbox{and for all subsets $Y$ of $\YY$ we have} \\[1ex]
        \quad \Pr(\,M(q(b_1))\,{\in}\,Y\,) \Wide{\leq} e^{\Vep}\DM\Pr(\,M(q(b_2))\,{\in}\,Y) \Q.
    \end{array}
\EE

With the above metric-based point of view we can say that an \VDP\ mechanism is (simply) a \VEP-Lipschitz function from databases $\DB$ with metric $\Dd$ to distributions of observations $\Dist\YY$ with metric $\DD$ \cite{Chatzi2013}.

\begin{definition}\label{d1048} (\DD/\Dd\ \VDP\ for mechanisms)\quad
A Lipschitz mechanism $M$ from \XX\ raw query outputs) to \YY\ (observations) is (\DD/\Dd) \VEP-Differentially Private just when
\BE{l1048}
    \begin{array}{l}
        \mbox{for all inputs $x_1,x_2$ in \XX\ to $M$ we have} \\
        \DD(M(x_1),M(x_2)) \Wide{\leq} \Vep\DM \Dd(x_1,x_2) \Q, \\
    \end{array}    
\EE
in which we elide \DD\ and \Dd\ when they are clear from context.
In \Eqn{e0944} we gave the special case where $M$'s inputs $x_1,x_2$ were raw query-results $q(b_1),q(b_2)$, i.e.\  with $b_1,b_2$ two databases acted on by the same query-function $q$. And \Eqn{e1057} was further specialised to  where the two databases were adjacent and the metric was $\Dd_H$, the Hamming distance.
\end{definition}

\SSection[s1114]{``Counting'' queries}

Counting queries on databases are the special case where the codomain $\XX$ of the query (the mechanism input) is the non-negative integers and the query $q$ returns the number of database rows satisfying some criterion, like ``being a man''. The ``average height'' query is not a counting query.

When the database metric is the Hamming distance $\Dd_H$, a counting query can be characterised more generally as one that is a 1-Lipschitz function \WRT\ $d_H$ and the usual metric (absolute difference) on the integers, i.e.\ one whose result changes by at most 1 between adjacent databases. Since composition of Lipschitz functions (merely) multiplies their Lipschitz factors, the composition of a counting query and an obfuscating mechanism is \VDP\ as a whole if the mechanism on its own (i.e.\ without the query, acting ``obliviously'' on $x{=}q(b)$) is \VEP-Lipschitz. That is why for counting queries we can concentrate on the mechanisms alone (whose type is $\XX{\Fun}\Dist\YY$) rather than including the databases and their type $\DB$ in our analysis.

\SSection[s1054]{Prior knowledge, open-source and the observer}
Although the database contents are not (generally) known, often the distribution of its query results \emph{is} known: this is ``prior knowledge'', where e.g.\ it is known that a database of heights in the Netherlands is likely to contain higher values than a similar database in other countries --- and that knowledge is different from the (unknown) fact we are trying to protect, i.e.\ whether a \emph{particular} person's height is in that database.

We abstract from prior knowledge of the database by concentrating instead on the prior knowledge $\pi$ of the distribution \XX\ of raw queries, the inputs $x$ to the mechanism, that is \emph{induced} as the push-forward of the query-function (an ``open source'' aggregating function) over the known distribution of possible databases themselves. Knowing $\pi$ on the input in \XX\, the observer can use her knowledge of the mechanism (also open source) to deduce a distribution on the output observations in \YY\ that will result from applying it and \mbox{--further--} she can also deduce a posterior distribution on \XX\ based on any particular $y$ in \YY\ that she observes.

\SSection{The Geometric mechanism is \VDP\ for $\Dd_H$}
\SSSection[s0915]{Specialising to $\Dd_H$}
Recall from \Sec{s1114} that $\Dd_H$, the Hamming distance, is what is typically used for counting queries. In that case we see as follows from \Eqn{e0944} that the Geometric mechanism $\Geo$ can be made \VDP. 

The Geometric \underline{distribution} centred on 0 with parameter $\alpha$ assigns (discrete) probability
\BE{e0959}
    \GeoD_\alpha(n) \WideDefs \NF{1-\alpha}{1+\alpha}\DM\alpha^{|n|}
\EE
to any integer $n$ (positive or negative) \cite{Ghosh:12}.
It implements an \VDP\ Geometric \underline{mechanism} by obfuscating the query according to \Eqn{e0959} above: thus set $\alpha\Defs e^{-\Vep}$ and define
\BE{e0917}
    \Geo^\Vep(n)(n') \Defs \GeoD_\alpha(n'{-}n) = \NF{1-\alpha}{1+\alpha}\DM\alpha^{|n'-n|}
\EE
to be the probability that integer $n$ is input and $n'$ is output. Thus applied to some $n$, the effect of $\Geo^\Vep$ with $\Vep\Defs -\ln\alpha$\,%
\footnote{The $\alpha$ in $G_\alpha$ is ${<}1$, so $\Vep{>}0$.}
is to leave $n$ as it is with probability
$\NF{1-\alpha}{1+\alpha}$
and to split the remaining probability $\NF{2\alpha}{1-\alpha}$ equally between adding 1's or subtracting them: $\Geo^\Vep$ continues (in the same direction) with repeated probability $\alpha$ until, with probability $1{-}\alpha$, it stops.

As explained in \Sec{s1054}, we now concentrate on $\Geo^\Vep$ alone and how it perturbs its input (a query result), i.e.\ no longer considering the database from which the query came.\,%
\footnote{Note that although the (raw) query is \emph{output} from the database, it is \emph{input} to the obfuscating mechanism. That is why we refer to \XX\ as ``input''.}


\SSSection[s1718]{The geometric mechanism truncated}
In \Eqn{e0917} the mechanism $\Geo^\Vep$ can effect arbitrary large perturbations. But in practice its output is constrained (in the discrete case) to a finite set $(0..N)$ by  (re-)assigning all probabilities for negative observations to observation 0, and all probabilities for ${\geq}N$ observations to $N$. For example with $e^{-\Vep}{=}\alpha{=}\NF{1}{2}$ and restricting to $(0..2)$ we have $\Geo^\Vep(0)(0)=\cdots{+}\NF{1}{12}{+}\NF{1}{6}{+}\NF{1}{3}=\NF{2}{3}$ and $\Geo^\Vep(0)(1)=\NF{1}{6}$ and $\Geo^\Vep(0)(2)=\NF{1}{12}{+}\NF{1}{24}{+}\cdots=\NF{1}{6}$.
It can be shown \cite{Chatzi:2019} however that truncation makes no difference to our results, and so from here on we will assume that truncation has been applied to $\Geo^\Vep$.

\SSection[s111359]{Discrete optimality}
It has been shown \cite{Ghosh:12} that when \XX\ is discrete
(and hence the prior $\pi$ on \XX\ is also), and when the obfuscation is via $\Geo^\Vep$, and when the observer applies a ``loss function'' $\ell(w,x)$ of her choice to monetise in $\RealNN$ the loss of utility to her if the raw query was $x$ but she assumes it was $w$, then \emph{any other} \VDP\ mechanism $M^\Vep$ acting on \XX\ can only lose \emph{more} utility (on average) according to that $\pi$ and $\ell$ than $\Geo^\Vep$ does. That is, the \VDP\ Geometric mechanism is \emph{optimal} for minimising loss (maximising utility) over all priors $\pi$ and all (``legal'') loss functions $\ell$ under a mandated \VDP\ obfuscation. A loss function is said to be \emph{legal} if it is monotone (increasing) wrt. to the difference between the guess ($w$) and the actual value $x$ of the query. As explained in \cite{Ghosh:12} this means that the loss $\ell(w, x)$ takes the form of a function $m(|w-x|, x)$, which must be monotone (increasing) in its first argument.

\SSection[s1104]{The geometric mechanism is never \VDP\ \\
on dense continuous inputs, e.g.\ when \Dd\ on \XX\ is Euclidean}

If the input metric for $\Geo$ is not the Hamming distance $\Dd_H$, e.g.\ when the \Geo's input \XX\ is continuous, still $G$'s output remains discrete, taking some number of steps, each of fixed length say $\lambda{>}0$, in either direction. 
That is, any $G$ input $x$ is perturbed to $x{+}i\lambda$ for some integer $i$.

Now because \XX\ is continuous and dense, we can vary the input $x$ itself, by a tiny amount, to some $x'$ so that $\Dd(x,x')\,{<}\,\lambda$ no matter how small $\lambda$ might be, producing perturbations
$x'{+}i\lambda$ each of which is distant that same (constant) $\Dd(x,x')$ from the original $x{+}i\lambda$ and, precisely because $\Dd(x,x')\,{<}\,\lambda$\,, those new perturbations cannot overlap the ones based on the  original $x$. 
%
%
%

Thus the two distributions produced by $G$ acting on $x$ and on $x'$ have supports that do not intersect at all. And therefore the $\DD$ distance between the two distributions is infinite, meaning that $G$ cannot be be \VDP\ for any (finite) \VEP.
That is, for a database producing truly real query results \XX, a standard (discrete) \Geo\ cannot establish \VDP\ for any \VEP, however large \VEP\ might be.

There are two possible solutions.
The first solution, both obvious and practical, is to ``discretise'' the input and to scale appropriately: a person's height of $1.75\textrm{m}$ would become $175\textrm{cm}$ instead.
A second solution however is motivated by taking a more theoretical approach.
Rather than discretise the type of the query results, we \emph{leave} it continuous --- and seek our optimal mechanism among those that --unlike the Geometric-- do not take only discrete steps. It will turn out to be the Laplace distribution.

\SSection{Our result --- continuous optimality}
In the discrete case typically the set \XX\ of raw queries is $(0..N)$ for some $N{\geq}0$, and the prior knowledge $\pi$ is a (discrete) distribution on that. For our continuous setting we will use $\XX{=}[0,1]$ for raw queries, the unit interval \UI, and the discrete distribution $\pi$ will become a proper measure on $[0,1]$ expressed as a probability density function. The \VDP\ obfuscating mechanisms, now $K^\Vep$ for ``kontinuous'', will take a raw query $x$ from a continuous set \XX\ rather than a discrete one.
And the metric on $\XX{=}\UIK$ will be Euclidean.

\textbf{Our (continuous) optimality result} formalised at \Thm{t1544} is that \VDP\ Laplace mechanism $\Lap^\Vep$ minimises loss  over all continuous priors $\pi$ on $\XX{=}\UIK$ and all legal loss functions $\ell$ under a mandated \VDP\ obfuscation with respect to the Euclidean metric on the continuous input $\XX{=}[0,1]$.

The theorem requires that \emph{all} mechanisms satisfy \Eqn{e0944} with $\Dd$ the Euclidean distance on continuous inputs. We write \VDP\ for such mechanisms. The argument in \Sec{s1104} above shows therefore that Geometric mechanisms are no longer suitable (for optimality) because on continuous \XX\ they are no longer \VDP.

\Section[s1459]{An outline of the proof}
We access the existing discrete results in \Sec{s1517}, and \Sec{s111359} from within the continuous \UI\ by ``pixelating'' it, that is  defining $\UI_N{=}\{0,\NF{1}{N},\NF{2}{N},\ldots,\NF{N-1}{N},1\}$ for integer $N{>}0$, and mapping $(0.N)$ isomorphically onto that discrete subset. We then establish near optimality for a similarly pixelated Laplace mechanism, showing that ``near'' becomes ``equal to'' when $N$ tends to infinity. In more detail:
\begin{enumerate}[(a)]

\item\label{i0938-602}(We show in \Sec{s1012} that) Any (discrete) prior on $\UI_N$ corresponds to some prior on the original $\UI$, but can also be obtained by pixelating some \emph{continuous} prior $\pi$ on all of \UI, concentrating its (now discrete) probabilities onto elements of $\UI_N$ only: e.g.\ the probability $\pi[\NF{n}{N},\NF{n+1}{N})$ of the entire $\NF{1}{N}$-sized interval is moved onto the point $\NF{n}{N}$.
We write it $\pi_N$.

\item\label{i0938-603}(\Sec{s0902}) Any function $f$ acting on all of \UI\ can be made into an \NStep\ function by first restricting its inputs to $\UI_N$ and then filling in the ``missing'' values $f(x)$ for $x$ in $(\NF{n}{N},\NF{n+1}{N})$ by copying the value for $f(\NF{n}{N})$. If $f$ is an \VDP\ mechanism $K^\Vep$, we write its \NStepped\ version as $K^\Vep_N$, and note that $K^\Vep_N$ remains \VDP\ when restricted to the points in $\UI_N$ only.

If $f$ is a loss function on (\WW\ and) \XX\ we write $\ell_N$ for its stepped version. 

\item\label{i0938-605} (\Sec{ss956}; \Lem{l1446}; \Lem{l1534-aa}) Now for any $N$, mechanism $K^\Vep_N$, prior $\pi_N$, and legal \NStep\ loss function $\ell_N$ we can appeal to the discrete optimality result: for the pixelated prior $\pi_N$ and the \NStep\ and legal $\ell_N$ the loss due to $\Geo^\Vep_N$ is $\leq$  the loss due to $K^\Vep_N$.

\item\label{i0938-606} (\Sec{ss957}; \Thm{t1431}) The replacement of $\Geo^\Vep_N$ by $\Lap^\Vep_N$ (both \NStep\ functions on $[0,1]$) is via pixelating the \emph{output} (continuous) distribution of $\Lap^\Vep_N$ to a multiple $T$ of $N$: we write that $^T\Lap^\Vep_N$. The Kantorovich-Rubinstein Theorem, provided additionally that $\ell_N$ is $p$-Lipschitz for some $p{>}0$ independent of $N$, shows that the (additive) difference between the $\Geo^\Vep_N$-loss and the $^T\Lap^\Vep_N$-loss, for any $\pi_N$ and $\ell_N$ and $T$ a multiple of $N$, tends to zero as $N$ increases.

\item\label{i0938-607} (\Sec{ss1410}) Then we remove the subscript $\pi_N$ on the prior, and on the mechanisms $K^\Vep_N$ and $^T\Lap^\Vep_N$, relying now on the \VDP\ of the two mechanisms to make the (multiplicative) ratio between the losses they cause tend to 1.

\item\label{i0938-608} (\Sec{s1352}) The final step, removing the subscript $_N$ from $\ell_N$, is that the loss-calculating procedure is continuous and that $\ell_N$ tends to $\ell$ as $N$ tends to infinity.

\end{enumerate}

\Section{Channels; loss functions; hyper-distributions; refinement}
In this section we provide a summary of the more general Quantitative Information Flow techniques that we will need for the subsequent development.
\SSection[s1144]{Channels, priors, marginals, posteriors}

The standard treatment of information flow is via Shannon's (unreliable) channels: they take an input $x$ from say $\XX$ and deliver an output that for a perfect channel will be $x$ again, but for an imperfect channel might be some other $x'$ in $\XX$ instead. For example, an imperfect channel transmitting bits might ``flip' input  bits so that with probability say $\NF{1}{4}$ an input 0 becomes an output 1 and vice versa \cite{Shannon:48}.
In the discrete case, and generalising to allow outputs of possibly a different type $\YY$, such channels are $\XX\Times\YY$ matrices $C$ whose row-$x$, column-$y$ element $C_{x,y}$ is the probability that input $x$ will produce output $y$. A perfect channel would be the identity matrix on $\XX{\Times}\XX$; a completely broken channel on $\XX{\Times}\YY$ for any $\YY$ would have $C_{x,y}= \NF{1}{\Size\YY}$.

The $x$-th row of a (channel) matrix $C$ is $C_{x,-}$; and the $y$-th column is $C_{-,y}$. Since each row sums to 1 (making $C$ a \emph{stochastic} matrix), the row $C_{x,-}$ determines a discrete distribution in $\Dist\YY$; for the ``broken'' channel it would be the uniform distribution, which we write \UNIF. 

As a matrix, a channel has type $\XX{\Times}\YY\,{\Fun}\,\Real$ (but with 1-summing rows); isomorphically it also has type $\XX{\Fun}\Dist\YY$. We'll write $\XX{\MFun}\YY$ for both, provided context makes it clear which one we are using.

If a \emph{prior} distribution $\pi\In\Dist\XX$ on $\XX$ is known, then the channel $C$ can be applied to $\pi$ to create a joint distribution $J$ in $\Dist(\XX\Times \YY)$ on both input and output together, written $\JD{\pi}{C}$ and where $J_{x,y}\,{\Defs}\;\pi_x C_{x,y}$. For that $J$, the left-marginal $\sum_y J_{x,y}$ gives the prior $\pi$ again (no matter what $C$ might be), i.e.\ the probability that the input was $x$ --- thus $\pi_x\,{=}\,J_{x,\Sigma}$ if we use that notation for the marginal.
The right marginal $J_{\Sigma,y}$ is the probability that the output is $y$, given both $\pi$ and $C$.

The $y$-\emph{posterior} distribution on \XX, given $\pi,C$ and a particular $y$, is the \emph{conditional} distribution on \XX\ if that $y$ was output: it is the $y$-th column divided by the marginal probability of that $y$, that is  $J_{-,y}/J_{\Sigma,y}$ (provided the marginal is not zero).

If we fix $\pi$ and $C$, and use the conventional abbreviation $p_{XY}$ for the resulting joint distribution $(\JD{\pi}{C})$, then the usual notations for the above are $p_X$ for left marginal $({=}\,\pi)$ and $p_X(x)$ for its value $\pi_x$ at a particular $x$, with $p_Y$ and $p_Y(y)$ similarly for the right marginal. Then $p_{X|y}(x)$ is the posterior probability of the original observation's being $x$ when $y$ has been observed. Further, we can write just $p(x)$ and $p(y)$ and $p(x|y)$ when context makes the (missing) subscripts clear.

\SSection[s0947]{Loss functions; remapping}
Our obfuscating mechanisms $M$ and $\Geo^\Vep$ are channels like $C$ in the discrete case --- the result of the query is the channel's input $x$, and the (perturbed) value the observer sees is the channel's output $y$.
The loss functions $\ell(w,x)$ will quantify the loss to her of seeing (only) $y$, and then choosing $w$, when what she really wants to know is $x$. Such \VDP\ mechanisms have earlier been modelled this way, i.e.\ as channels by Alvim et al.\ \cite{DBLP:conf/icalp/AlvimACP11} and Chatzikokolakis et al.\ \cite{Chatzi2013},
who observed that for \VDP\ the ratios of their entries must satisfy the \VDP\ constraints, because
the definition at \Sec{s125718}\Eqn{l1048} reduces to comparing (multiplicatively) adjacent entries in channel columns.

The connection between the observation $y$ and the loss-function parameter $w$ is that the observer does not necessarily have to ``take what she sees'' --- there might be good reasons for her making a different choice. For example, in a word-guessing game where the last, obfuscated letter {\tt?} in a word \framebox{\tt SA?} is shown on the board, the observer might have to guess what it really is. Even if it looks like a blurry \texttt{Y} (value 4 in Scrabble), she might instead guess \texttt{X} (value 8) because that would earn more points on average if from prior knowledge she knows that {\tt X}  strictly \emph{more} than half as likely as {\tt Y} is --- i.e.\ it's worth her taking the risk. Thus rather than mandating that the observer must accept what she thinks the letter is most likely to be, she uses the obfuscated query $y$ to deduce information about the \emph{whole} posterior distribution of the \emph{actual} query\ldots\ and might suggest that she guess some $w{\neq}y$, because the expected loss of doing that is less than (the expected utility is greater than) it would be if she simply accepted the $y$ she saw. That rational strategy is called ``remapping'' \cite{Ghosh:12}. Thus she sees $y$, but $y$ tells her that $w$ is what she should choose as her least-loss inducing guess for $x$. That is, the \emph{simplest} strategy is ``take what you see''; but it might not be the best one. In general (and now using $M$ again for mechanism), we write $\Loss{\pi}{M}{\ell}$ for the \emph{expected} loss to a rational observer, given the $\pi,M$ she knows and the loss function $\ell$ she has chosen: it is
\BE{e1535}
\sum_y p(y)~~ \Min_w \sum_x \ell(w,x)\,p(x|y) \Q,
\EE
that is the expected value, over all possible observations $y$ and their marginal probabilities, of the \emph{least} loss she could rationally achieve over \emph{all} her possible choices $w$ given the knowledge that $y$ will have provided about the posterior distribution $p(X|y)$ of the actual raw input $x$.
Note that $M$ and $\pi$ determine (from (\Sec{s1144}) the $p(y)$ and $p(x|y)$ that appear in \Eqn{e1535}.  We remark that this formulation for measuring expected loss corresponds precisely to the formulation used by Ghosh et al.\ in the optimality theorem. 

\SSection{The relevance of hyper-distributions, abstract channels}

It is important to remember that
the expected-loss formula \Eqn{e1535} does not use the \emph{actual} mechanism-output values $y$ in any way directly: instead it takes the only expected value of what they \emph{might be}. All that matters is their marginal probabilities $p(y)$ and the a-posteriori distributions $p(X|y)$ that they induce.
That allows us to abstract from \YY\ altogether.

A \emph{hyper-distribution} expresses that abstraction: it is a distribution of distributions on \XX\ alone, that is of type $\Dist\Dist \XX$;
abbreviate those as ``hyper'' and ``$\Dist^2\XX$''.
Given a joint distribution $J\In\Dist(\XX{\Times}\YY)$, we write $\Hyp{J}$ for the hyper-distribution whose support is posterior distributions\,%
\footnote{In the hyper-distribution literature these are called ``inners'' \cite{Alvim:20a}.}
$p(X|y)$ on $X$ and which assigns the corresponding marginal distribution $p(y)$ to each. (Zero-valued marginals are left out.) We now re-express \Eqn{e1535} in those terms.

If we write $\ell(w,-)$ for the function on \XX\ that $\ell$ determines once $w$ is fixed, and write $\,\Exp{\textsc {\tiny dist}}{\textsc{\footnotesize rv}}\,$ for expected value of random-variable \textsc{\small rv} with distribution \textsc{\small dist}, then $\Min_w\,\Exp{p(X|y)}{\ell(w,-)}$ is the inner part of \Eqn{e1535}. Then fix some $\ell$ and define for general distribution $\delta\In\Dist X$ that 
\BE{e1624}
	Y_\ell(\delta)\WideDefs\Min_w\,\Exp{\delta}{\ell(w,-)} \Q, 
\EE
(using $Y$ for ``entrop$Y$'') so that  $Y_\ell$ is itself a real-valued function on distributions $\delta$ (as e.g.\ Shannon entropy is). With that preparation, the expression \Eqn{e1535} becomes the expected value of $Y_\ell$ over the hyper produced by abstracting from $J=\JD{\pi}{M}$ as above. That is \Eqn{e1535} gives equivalently
\BE{e1536}
\Loss{\pi}{M}{\ell} \Wide{=}
\Exp{\JDH{\pi}{M}}{Y_\ell} \Q,
\EE%
in which the $M$ and $\pi$ now explicitly appear and where
\mbox{--we} recall-- the brackets $\Hyp{-}$ convert the joint distribution $\JD{\pi}{M}$ to a hyper.
(If $Y_\ell$ were in fact Shannon entropy, then \Eqn{e1536} would be the \emph{conditional} Shannon entropy. But $Y_\ell$'s are much more general than Shannon entropy alone \cite{Alvim:2014aa,Alvim:2012aa}.)

Finally, using hypers we define an \emph{abstract} channel to be a function from prior to hyper, i.e.\ of type $\XX{\Fun}\Dist^2\XX$, realised from some concrete channel $M\In\XX{\MFun}\YY$ as $\pi\,{\mapsto}\,\JDH{\pi}{M}$. It is ``abstract'' because the type \YY\ no longer appears: it is unnecessary because if $M(\pi)$ is the application of $M$ as a function applied to prior $\pi$, then from \Eqn{e1536} the worst rational expected loss is written simply $\Exp{M(\pi)}{Y_\ell}$~.

(Recall from \Sec{s0947} that this naturally takes into account the ``rational observers'' and the remapping they might perform, as described in \cite{Ghosh:12}.)

\SSSection{Example of a channel representation of a mechanism}

If we have a discrete input $\XX{\Defs}\{x_0, x_1, x_2 \}$, and discrete output $\YY{\Defs}\{y_0, y_1, y_2, y_3, y_4\}$, we can represent an obfuscating mechanism $M$ with the channel $M$ below.

\vspace{2mm}
  \[
 M = 
    \begin{bmatrix}
     \nicefrac{2}{3} & \nicefrac{1}{6} & \nicefrac{1}{12} & \nicefrac{1}{24} & \nicefrac{1}{24} \\
     \nicefrac{1}{6} & \nicefrac{1}{6} & \nicefrac{1}{3} & \nicefrac{1}{6} & \nicefrac{1}{6} \\
     \nicefrac{1}{24} & \nicefrac{1}{24} & \nicefrac{1}{12} & \nicefrac{1}{6} & \nicefrac{2}{3}
  \end{bmatrix} 
  \]
  
 As described in \Sec{s1144} above, the row $M_{x,-}$ corresponds to the probability distribution of outputs $y$ in $\YY$ for that $x$. For example the top left number $  \nicefrac{2}{3}{=}M_{x_0, y_0}$ is the probability that output $y_0$ is observed when the input is $x_0$. We can interpret this as an \VDP\ mechanism once we know the metric $\Dd$ on $\XX$. In particular \Sec{s125718}\Eqn{e0933} simplifies to comparing ratios of entries in the same column, and when we do that we find that for example $\DD(M(x_0), M(x_1))= \ln 4$. Thus from \Sec{s125718}\Eqn{l1048} now applied to $\XX$ we can say that if $M$ is \VDP\ then $\Vep$ satisfies
 \[
 \DD(M(x_0), M(x_1))  \Wide{=} \ln 4 \Wide{\leq}  \Vep\DM \Dd(x_0, x_1) \Q.
 \]

\SSSection{Example of a loss function calculation}

Now suppose that we choose a  loss function known as ``Bayes Risk'',  ${\sf br}$ defined on $\XX{\Defs}\{x_0, x_1, x_2\}$ as above:
\[
{\sf br}(w, x)\Wide{\Defs} 1 ~\textrm{~if~}~ x\neq w~ \textrm{~else~}~ 0~,
\]
where $\WW{\Defs}\XX$. Letting the input prior be the uniform distribution $\UNIF$ over $\XX$, we can compute the loss $\Loss{\UNIF}{M}{{\sf br}}$ by selecting for each output $y$, the $w$ which makes the expected value of ${\sf br}(w, -)$ over the posterior  $p_{X|y}$ the least. We then take the expected value of these least values over the marginal $p_Y$. For $y_0$ for example, that least expected value occurs at $w{=}x_0$, and for $y_1$ it occurs \emph{either} for $w{=}x_0$ or $w{=}x_0$. Overall the total expected loss is $\nicefrac{1}{3}$. 

\SSection[s0931]{Refinement of hypers and mechanisms}
The hypers $\Dist^2\XX$ on $\XX$ have a partial order $(\Ref)$ ``refinement'' \cite{McIver:13} that we will need in the proof of our main result. It admits several equivalent interpretations in this context. Below, we write $\Delta$ etc.\ for general hypers in $\Dist^2\XX$.

We have that $\Delta\,{\Ref}\,\Delta'$, that hyper $\Delta$ is refined by hyper $\Delta'$, under any of these equivalent conditions:
\begin{enumerate}[(a)]
\item\label{i1327-a} when $\Exp{\Delta}{Y_\ell} \leq \Exp{\Delta'}{Y_\ell}$ for \emph{all} loss functions $\ell$ (i.e.\ whether legal or not).
\item\label{i1327-b} when considered as distributions on posteriors $\Dist\XX$ it is possible to convert $\Delta$ into $\Delta'$ via a Wasserstein-style
``earth move'' of probability from one posterior to another \cite{Rachev:98,Deng:09,Alvim:20a}.

\item\label{i1327-c} when generated from joint-distribution matrices $D$ in $\Dist(\XX{\times}\YY)$ generating $\Delta$, and $D'$ in $\Dist(\XX{\times}\YY')$ generating $\Delta'$, there is a ``post-processing matrix'' $R$ of type $\YY{\MFun}\YY'$ such that as matrices we have $D{\cdot}R = D'$ via matrix multiplication.
\end{enumerate}
And we say that one mechanism $M$ is refined by another $M'$ just when $\JDH{\pi}{M}\,{\Ref}\,\JDH{\pi}{M'}$ for all priors $\pi$. When this occurs we also write $M\Ref M'$.
From formulation \Itm{i1327-a} we will use the fact that the $(\Ref)$-infimum of the $^T\Lap^\VEP_N$'s (indexed over a sequence of $T$'s)
is just $\Lap^\Vep$ itself \cite{McIver:12} and \cite[Lem. 20, Appendix {\textsection}B]{Optimality:FullPaper:2021}.

Formulation \Itm{i1327-b} is particularly useful. If we find a specific earth move from $\Delta$ to $\Delta'$ that defines a refinement we can then use the equivalent \Itm{i1327-a} to deduce that $\Exp{\Delta}{Y_\ell} \leq \Exp{\Delta'}{Y_\ell}$. However if we can also compute the cost \footnote{The cost is determined by the amount of ``earth'' to be moved, and the distance it must be moved. See for example \cite{Lawler:76}.} of the particular earth move we can conclude in addition that the difference $|\Exp{\Delta}{Y_\ell} - \Exp{\Delta'}{Y_\ell}|$ must be bounded above by an amount we can compute. This follows from the well-known Kantorovich-Rubinstein duality \cite{Rachev:98} which says that $|\Exp{\Delta}{Y_\ell} - \Exp{\Delta'}{Y_\ell}|$ is no more than minimal cost incurred by any earth move transforming $\Delta$ to $\Delta'$ scaled by the ``Lipschitz constant'' \footnote{The Lipschitz constant of a function is the amount by which the difference in outputs can vary when compared to the difference in inputs.} of $Y_\ell$. We use these ideas in \Lem{l1218} and \Thm{t1431}.

\Section{Measures on continuous \XX\ and \YY}\label{s1732}

\SSection{Measures via probability density functions}

Continuous analogues of the $\pi$, $M$ and $\ell$ will be our principal concern here: ultimately we we will use $\Meas[0,1]$ for our measurable spaces \XX\ and \YY, and will suppose for simplicity that $\XX{=}\YY{=}[0,1]$. (More generality is achieved by simple scaling).

Measures $\Meas[0,1]$ (that is $\Meas\XX$ and $\Meas\YY$) will be given as probability density functions, where a \PDF\ say $\mu\In[0,1]{\Fun}\RealNN$ determines the probability $\int^b_a \mu$ assigned to the sample $[a,b]{\subseteq}[0,1]$ using the standard Borel measure on $[0,1]$, and more generally the expected value of some random variable $V$ on $[a,b)$ given by \PDF\ $\mu$ is  $\int_a^{b-}\mu(x)V(x) \DX x$\,.\,%

Even though $\mu$ is of type \PDF, we abuse notation to write for example $\mu[a,b)$ for the probability $\int_a^{b-}\kern-.75em\mu$ that $\mu$ assigns to that interval, and $\mu_a$ for the probability $\mu$ assigns to the point $a$ alone, i.e.\ some $r$ just when when the actual \PDF-value of $\mu(a)$ is the Dirac delta-function scaled by $r$, written  $\Dirac{r}$\,.

\SSection{Continuous mechanisms over continuous priors}
Our mechanisms $M$, up to now discrete, will now become ``kontinuous'', renamed $K$ as a mnemonic. Thus
a continuous mechanism $K\In \XX{\Fun}\Meas\YY$ given
input $x$ produces measure $K(x)$ on the observations $\YY{=}[0,1]$.
And given a a whole continuous prior $\pi\In\Meas[0,1]$, that same $K$ therefore determines a joint measure over $\XX{\times}\YY$.\,%
\footnote{See \cite[Appendix {\textsection}A2]{Optimality:FullPaper:2021}.}
By analogy with (\ref{e1624},\ref{e1536}) we have
\begin{definition}\label{d1422} \textbf{Continuous version of \Eqn{e1535}}\quad
The expected loss $\Loss{\pi}{K}{\ell}$ due to continuous prior $\pi$, continuous mechanism $K$ and loss function $\ell$ is given by \footnote{This is well defined whenever the $\WW$-indexed family  of functions of $y$ given by $\kern-.2em\int^1_0 \ell(w,x)\pi(x)K(x)(y)~\DX x$ contains a countable subset ${\cal W}'$ such that the  $\Inf$ over $\WW$ is equal to the $\Inf$ over  $\WW'$ \cite{Meyer-Nieberg:91}. This is clear if $\WW$ is finite, and whenever $\WW'$ can be taken to be the rationals.}
\BE{e1637}
	\int^1_0 (\,\Inf_w~(\kern-.2em\int^1_0 \ell(w,x)\pi(x)K(x)(y)~\DX x)\,)\DX y \,.
\EE
\end{definition}
The continuous version of uncertainty \Eqn{e1624} is now
\[
    Y_\ell(\delta) \WideDefs \Inf_{w\In\WW}\int_0^1 \kern-.3em\ell(w, x)\delta(x) \DX x
\]
and the continuous version of expected loss \Eqn{e1536} is now
\[
\Loss{\pi}{K}{\ell} \Wide{=}
\int_{y\In\YY} \kern-.5emY_\ell\,\DX {\mbox{\small$K(\pi)$}} \Q.
\]

\SSection[s160734]{The truncated Laplace mechanism}
As for the Geometric mechanism, the Laplace \underline{mechanism} is based on the Laplace \underline{distribution}. It is defined as follows:
\begin{definition}{(Laplace distribution)}\label{d1432}
The \emph{$\varepsilon$-Laplace mechanism} with with input $x$ in $\XX{=}[0,1]$ and probability density in $\RealNN$ for output $y$ in \YY\ is usually written as a  \PDF\ in $y$ (for given $x$) as
\cite{Wilson:1923}
\[
	\LV(x,y) \WideDefs   \varepsilon/2\cdot e^{-\varepsilon |y{-}x|} \Q.
\]
The $\varepsilon/2$ is a normalising factor.
It is known \cite{Chatzi2013} that the mechanism $\LV$ satisfies \VDP\ over $[0,1]$ (where the underlying metric on $\XX$\ is Euclidean). Just as for the Geometric mechanism we truncate $\LV$'s outputs so that they also lie inside $\UI$. We do so in the same manner, by remapping all outputs greater than $1$ to $1$, and all outputs less than $0$ to $0$.
\end{definition}

\begin{definition}{(truncated Laplace mechanism)}\label{d1231}
As earlier for $\Geo^\Vep$, we \emph{truncate} the Laplace mechanism \LV\ to
$\LTV$ for inputs restricted to $[0,1]$, and output restricted to $[0,1]$, in the following way (as a \PDF):
\[
\begin{array}{llll}
\LTV(x)(y) &\Defs
    & \Dirac{a} & \textit{if $y{=}0$} \\
  & & \LV(x,y) & \textit{if $0{<}y{<}1$} \\
  & & \Dirac{b} & \textit{if $y{=}1$} \Q,
\end{array}
\]
where the constants $a,b$ are $\int_{-\infty}^0 \LV(x,y) \DX y = e^{\varepsilon x}/2$ and $\int_1^\infty \LV(x,y) \DX y = e^{\varepsilon (1-x)}/2$ respectively, and $\Dirac{r}$ is the Dirac delta-function with weight $r$.

\end{definition} 

We can now state our {\bf principal contribution}. It is to show that the \emph{truncated} Laplace $\LTV$ is universally optimal, in this continuous setting, in the same way that $\Geo^\Vep$ was optimal in the discrete setting:

\noindent\begin{theorem}\label{t1544} \textbf{(truncated Laplace is optimal)}
Let $K^\Vep$ be any continuous \VDP\ mechanism with input and output both $[0,1]$, and let $\pi$ be any continuous (prior) probability distribution over $[0,1]$ and $\ell$ any Lipschitz  continuous \footnote{Lipschitz continuous is less general than continuous. It means that the difference in outputs is within a constant $\kappa{>}0$ scaling factor of the difference between the inputs.}, legal loss function on $\XX{=}\UIK$.
\par
Then $\Loss{\pi}{\LTV}{\ell} \leq \Loss{\pi}{K}{\ell}$~.
\end{theorem}

\bigskip
As we foreshadowed in the proof outline in \Sec{s1459}, \Thm{t1544} relies ultimately on the earlier-proven optimality $\Geo^\Vep$ in the discrete case: we must show how we can approximate continuous \VDP\  mechanisms in discrete form, each one satisfying the conditions under which the earlier result applies, and in \Sec{s1447} we fill in the details. Along the way we show how the Laplace mechanism provides a smooth approximation to the Geometric- with discrete inputs. 

\Section[s1447]{Approximating Continuity for \XX}

\SSection{Connecting continuous and discrete}
Our principal tool for connecting the discrete and continuous settings is the evenly-spaced discrete subset $\UI_N=\{0,\NF{1}{N},\NF{2}{N}\ldots, \NF{N-1}{N},1\}$ of the unit interval $\UI{=}[0,1]$ for ever-increasing $N{>}0$.

The separation $\NF{1}{N}$ is the \emph{interval width}.

\SSection[s1012]{Approximations of continuous priors}
The $N$-approximation of prior $\pi\In\Meas\UIK$ of type $\Dist\UIK_N$, i.e.\ yielding actual probabilities (not densities), and
is defined
\[
	\pi_N(n/N) \WideDefs
		\begin{array}[t]{lp{10em}}
			\pi[\NF{n}{N},\NF{n+1}{N}) & if $n\,{<}\,N{-}1$ \\
			\pi[\NF{n}{N},1]         & if $n\,{=}\,N{-}1$ \\
			0                          & otherwise \Q.
		\end{array}
\]
The discrete $\pi_N$ gathers each of the continuous $\pi$-interval's measure onto its left point, with as a special case [1,1] from $\pi$ included onto the point $\NF{N-1}{N}$ of $\pi_N$.

As an example take $N$ to be $2$, and  $\pi$ to be the uniform (continuous) distribution over $\UI$, which can be represented by the constant $1$ \PDF. Since the interval width is $\nicefrac{1}{2}$, we see that $\pi_N$ assigns probability $\nicefrac{1}{2}$ to both $0$ and $\nicefrac{1}{2}$ and zero to all other points in $\UI$.

\SSection[s0902]{\NStep\ mechanisms and loss functions}
In the other direction,
we can lift discrete $M$ and loss-function $\ell$ on $\UI_N$ into the continuous $\XX{=}\UIK$ by replicating their values for the $x$'s \emph{not} in $\UIK_N$\, in a way that constructs \NStep\ functions: we have
\begin{definition}
For $x$ in $\UI{=}[0,1]$ define $\NFloor{x}{N} \Defs \Floor{Nx}/N$.
\end{definition}

\begin{definition}\label{d0945}
Given mechanism $M\In\UIK_N{\MFun\YY}$, define $M_N\In[0,1]{\Fun}\RealNN$ so that 
\[
	M_N(x) \WideDefs
		\begin{array}[t]{lp{10em}}
			M(\NFloor{x}{N}) & if $0{\leq} x{<}1$ \\
			M(\NF{N-1}{N}) & if $x{=}1$ \Q.
		\end{array}
\]
Note that we have not yet committed here to whether $M$ produces discrete or continuous distributions on its \emph{output} $\YY$. We are concentrating only on its input (from \XX).

Similarly, given loss function $\ell\In\WW{\times}\UI_N,{\Fun}\,\RealNN$, define $\ell_N\In\WW{\times}[0,1]\,{\Fun}\,\RealNN$ so that
\[
	\ell_N(w,x) \WideDefs
		\begin{array}[t]{lp{10em}}
			\ell(w,\NFloor{x}{N}) & if $0{\leq} x{<}1$ \\
			\ell(w,\NF{N-1}{N}) & if $x{=}1$ \Q.
		\end{array}
\]
Say that mechanisms and loss functions over $[0,1]$ are \emph{\NStep\ functions} just when they are constructed as above.
\end{definition}

The important property enabled by the above definitions is the correspondence between loss functions' values in their pixelated and original versions, which will allow us to apply the earlier discrete-optimality result, based on \Lem{l1720-a} to come.
That is, we have
\begin{lemma}\label{l1430-a}
For any continuous prior $\pi$ in $\Meas\UIK$, mechanism $M$ in $\UI{\MFun}\YY$
and loss function $\ell$ in $\WW{\times}\UI{\Fun}\RealNN$ we have
\[
	\overbrace{\raisebox{0pt}[2.5ex]{$\Loss{\pi}{M_N}{\ell_N}$}}^\textit{\small continuous \XX}
	\Wide{=}
	\overbrace{\raisebox{0pt}[2.5ex]{$\Loss{\pi_N}{M}{\ell}$}}^\textit{\small discrete \XX} \Q.
\]
That is, the loss realised via a pixelated $\pi_N$, and (already discrete) $M$ and $\ell$, all operating on $\UI_N$, is the same as the loss realised via the original continuous $\pi$ and the lifted (and thus \NStep) mechanism $M_N$ and $\ell_N$, now operating over all of $\XX{=}\UIK$.
\begin{IEEEproof}
We interpret the losses using \Def{d1422}, focussing on the integrand of the inner integral. Note that we can split it up into a finite sum of integrals of the form $\int_{\nicefrac{n}{N}}^{\nicefrac{n{+}1}{N}-}\pi(x)V(x) \DX x$. When we do that for the     left-hand formula $\Loss{\pi}{M_N}{\ell_N}$ we see that throughout the interval  $[\nicefrac{n}{N}, \nicefrac{n{+}1}{N})$ the contribution of the mechanism and the loss is constant, i.e.\
 $M_N(x)(y)\,{\DM}\,\ell_N(w,x) = M(\nicefrac{n}{N})(y)\,{\DM}\,\ell(w,\nicefrac{n}{N})$. This means the integral becomes 
 \[
  M(\nicefrac{n}{N})(y)\,{\DM}\,\ell(w,\nicefrac{n}{N})\,{\DM}\, \int_{\nicefrac{n}{N}}^{\nicefrac{n{+}1}{N}-}\pi(x) \DX x
 \]
 which is equal to $ M(\nicefrac{n}{N})(y)\,{\DM}\,\ell(w,\nicefrac{n}{N})\,{\DM}\, \pi_N(n/N)$. A similar argument applies to the last interval (which includes $1$), compensated for by the definitions of $\ell_N$ and $M_N$  to take their corresponding values from $\nicefrac{{1{-}N}}{N}$.
 
Looking now at the right-hand formula, $\Loss{\pi_N}{M}{\ell}$ we see that it is now exactly the finite sum of the integrals just described.
\end{IEEEproof}

\end{lemma}

\subsection{Approximating continuous \VDP\ mechanisms}\label{ss1410}
The techniques above give good discrete approximations for continuous-input \VDP\ mechanisms $M$ acting on continuous priors simply by considering $M_N$'s for increasing $N$'s, using \Sec{s0902}.
As a convenient abuse of notation, when we \emph{start} with a continuous-input mechanism $M$ on $[0,1]$ we write $M_N$ to mean the \NStep\ mechanism that is made by first restricting $M$ to the subset $\UI_N$ of $[0,1]$ and then lifting that restriction ``back again'' as in \Def{d0945}, effectively converting it into an \NStep\ function. When we do this we find that the posterior loss wrt.\ \NStep\ loss functions can be bounded above and below
by using pixelated priors and \NStepped\ mechanisms.

\begin{lemma}\label{l1720-a}
Let $K$ be a continuous-input \VDP\ mechanism, and $\pi$ in $\Meas[0,1]$ a continuous prior and $\ell$ a (non-negative) \NStep\ function. Then the following inequalities hold:
\[
       e^{-\frac{\Vep}{N}}\cdot\,
       \overbrace{\raisebox{0pt}[2.5ex]{$\Loss{\pi_N}{K_N}{\ell}$}}^{\it discrete \XX}
~\leq~ \overbrace{\raisebox{0pt}[2.5ex]{$\Loss{\pi}{K}{\ell}$}}^{\it continuous \XX}
~\leq~~e^{\frac{\Vep}{N}}\cdot\,
       \overbrace{\raisebox{0pt}[2.5ex]{$\Loss{\pi_N}{K_N}{\ell}$}}^{\it discrete \XX} \,.
\]

(Notice that the middle formula $\Loss{\pi}{K}{\ell}$, the mechanism $K$ is not \NStepped, but in the formulae on either side they are as in \Lem{l1430-a}.)
\begin{IEEEproof}
The proof is as for \Lem{l1430-a}, but noting also that $K$'s being \VDP\ implies that for all $N$ we have
 $K(\NFloor{x}{N})(y){\times}e^{-\frac{\Vep}{N}} \leq K(x)(y)\leq  K(\NFloor{x}{N})(y){\times}e^{\frac{\Vep}{N}}$.
 \footnote{Here we are using the \VDP-constraints applied to the \PDF\ $K(x)(y)$.}
\end{IEEEproof}
\end{lemma}

With \Lem{l1430-a} and \Lem{l1720-a} we can study optimality of \LTV\ 
on finite discrete inputs $\UI_N$. We will see that, although Geometric mechanisms are still optimal for the (effectively) discrete inputs $\UI_N$, the Laplace mechanism provides increasingly good  \emph{approximate optimality} for $\UI_N$ as $N$ increases, and is in fact (truly) optimal in the limit.

\Section[s0918]{The Laplace and Geometric mechanisms}
In this section we make precise the restriction
of the Geometric mechanism $G^\Vep$ (\Sec{s0915}) to inputs and outputs both in $\UI_N$ (a subset of $[0,1]$): for both $x,y$ in $\UI_N$ we define
\begin{equation}\label{e1327}
\overbrace{\raisebox{0pt}[2.5ex]{$\Geo_N^\Vep(x)(y)$}}^\textrm{\small on $\UI_N$}
\Wide{\Defs}
\overbrace{\raisebox{0pt}[2.5ex]{$\Geo^{\frac{\Vep}{N}}(Nx)(Ny)$}}^\textrm{\small on $(0..N)$} \Q.
\end{equation}

As an illustration, we take  $\Vep{=}2\ln4$ and input $\XX{=}\UIK_2$,
in which the 2 comes from $\UI_2$ and the $\ln4$ comes from the $\alpha{=}\NF{1}{4}$ of the Geometric distribution used to make the mechanism $\Geo^\Vep$.
Using the \emph{three} points $0, \nicefrac{1}{2}$ and $1$ of the input, we compute the truncated geometric mechanism $\Geo_2^\Vep$ as the channel
below, where the rows' labels are (invisibly) the inputs $\UI_2$, and the columns are similarly labelled by the outputs (also $\UI_2$ in this case). This means that if the input was $0$, then the output (after truncation) will be $0$ with probability $\NF{4}{5}$, and $1/2$ with probability $\NF{3}{20}$ etc:
\[
 \Geo_2^\Vep = 
  \begin{bmatrix}
     \nicefrac{4}{5} & \nicefrac{3}{20} & \nicefrac{1}{20} \\
     \nicefrac{1}{5} & \nicefrac{3}{5} & \nicefrac{1}{5} \\
     \nicefrac{1}{20} & \nicefrac{3}{20} & \nicefrac{4}{5} 
  \end{bmatrix} \Q.
  \]
Notice now that the ratio of adjacent probabilities that are in the same column satisfy the \VDP\ constraint, so for example $\nicefrac{4}{5}\div\nicefrac{1}{5} = \nicefrac{3}{5}\div \nicefrac{3}{20} = 4 \leq e^{(2\ln 4)/2} $. Notice also that the
distance between adjacent inputs in $\UI_2$ under the Euclidean distance is $\NF{1}{2}$, not 1 as it would be in the conventional $\XX{=}(0,1,2)$.
  
Suppose now that we consider $\UI_4$ instead, consisting of the \emph{five} points $0,  \nicefrac{1}{4},  \nicefrac{1}{2}, \nicefrac{3}{4}$ and $1$, and we adjust the $\alpha$ in the underlying Geometric \underline{distribution} $G_\alpha$ from \Sec{s0915}\Eqn{e0959}. The \VDP\ parameter \VEP, now $4\ln2$, is the \emph{same} as before
--- and the resulting matrix is
\[
  \Geo_4^\Vep = 
    \begin{bmatrix}
     \nicefrac{2}{3} & \nicefrac{1}{6} & \nicefrac{1}{12} & \nicefrac{1}{24} & \nicefrac{1}{24} \\
     \nicefrac{1}{3} & \nicefrac{1}{3} & \nicefrac{1}{6} & \nicefrac{1}{12} & \nicefrac{1}{12} \\
     \nicefrac{1}{6} & \nicefrac{1}{6} & \nicefrac{1}{3} & \nicefrac{1}{6} & \nicefrac{1}{6} \\
     \nicefrac{1}{12} & \nicefrac{1}{12} & \nicefrac{1}{6} & \nicefrac{1}{3} & \nicefrac{1}{3} \\
     \nicefrac{1}{24} & \nicefrac{1}{24} & \nicefrac{1}{12} & \nicefrac{1}{6} & \nicefrac{2}{3}
  \end{bmatrix} 
\]
  
 As before though, the ratio of adjacent probabilities that are in the same column satisfy the \VDP-constraint over all of $\UI_4$: now we have $\nicefrac{2}{3}\div\nicefrac{1}{3}=\nicefrac{1}{3}\div\nicefrac{1}{2} = 2 \leq e^{(4\ln 2)/4}$.
 
This amplifies the explanation in \Eqn{e0944} that the \VDP\ constraints over discrete inputs $\UI_N$ must take into account the underlying metric on the input space. More generally, whenever we double $N$ in $\UI_N$, the $\alpha$-parameter must become $\sqrt{\alpha}$.

At this point, we have enough to be able to appeal to the discrete optimality result, to bound below the losses for continuous mechanisms, provided that the loss $\ell_N$ is $N$-legal, i.e.\ that its legality obtains at least for the distinct points in $\UI_N$.

\begin{lemma}\label{l1446}
For any continuous prior $\pi$ in $\Meas\UIK$, \VDP-mechanism $M\In\UI{\MFun}\YY$ and  loss function $\ell\In\WW{\times}\UI{\Fun}\RealNN$ such that $\ell_N$ is $N$-legal, we have:
\[
	\Loss{\pi_N}{\Geo_N^\Vep}{\ell_N}
	\Wide{\leq}
	\Loss{\pi}{M_N}{\ell_N}
\]
\begin{IEEEproof}
Follows from \Lem{l1430-a} and noting that $M$ restricted to $\UI_N$ satisfies the conditions for universal discrete optimality \cite{Ghosh:12}.
\end{IEEEproof}
\end{lemma}

Our next task is to study the relationship between the Geometric- and Laplace mechanisms. We show first that $\Geo_N^\Vep$ is refined (\Sec{s0931}) by the truncated Laplace mechanism also restricted to to $\UI_N$.  Since $\LTV$ is already defined over the whole of $\UI$ we continue to write its restriction to $\UI_N$ as $\LTV$. This will immediately show that losses under the Geometric are no more than those under the Laplace (\Sec{s0931}(1)), consistent with observations that, on discrete inputs, Laplace obfuscation does not necessarily minimise the loss. Since the output \YY\ of $\LTV$ is continuous, we proceed by first approximating it using post-processing to make Laplace-based mechanisms $^T\LTV$, defined below,  which have discrete output, and which can form an anti-refinement chain converging to $\LTV$. We are then able to show separately the refinements between  $\Geo_N^\Vep$ and $^T\LTV$, using methods designed for finite mechanisms.

\begin{figure}
   \begin{subfigure}[b]{.9\columnwidth}
      \centering
      \includegraphics[width=.9\linewidth]{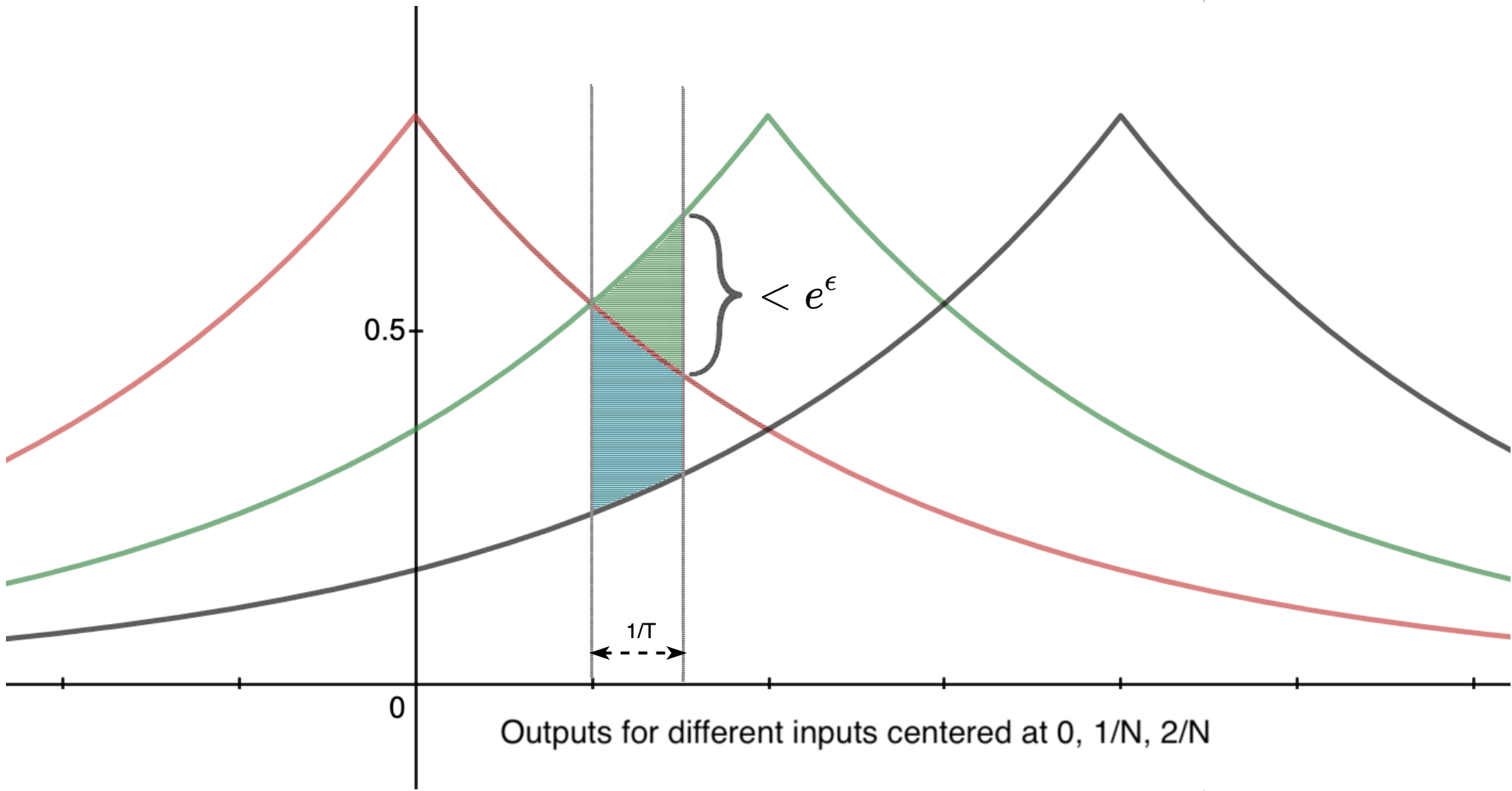}\\
      \medskip{\small The width of the central ``vertical slice'' is $1/T$.}
      \caption{Illustrates batching the output for $^T\Lap$ (similar for $^T\LTV$). The outputs (shown here as \PDF\ plots)  are batched into output segments of length $\nicefrac{1}{T}$ in this example, for $T{=}8$.  The segment from $[x, x{+}\nicefrac{1}{T})$ is indicated by the two vertical lines. The probability assigned to this segment is the area under the relevant curves. For  the red curve it is the sum of the white and blue regions; the green curve it is the sum of the white, blue and green regions and  for the black curve it is only the white region.}

\label{f1343a}
    \end{subfigure}
    
  \vspace*{8pt}%

    \begin{subfigure}[b]{.9\columnwidth}
      \centering
      \includegraphics[width=.9\linewidth]{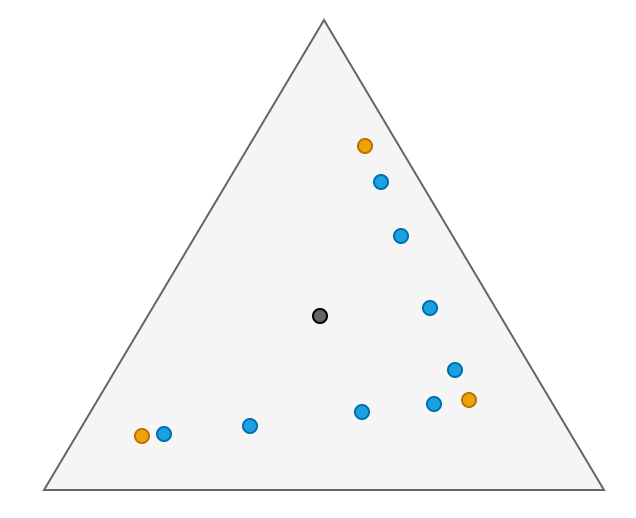}
      \caption{The supports of hypers $\JDH{\UNIF}{{\Geo_2^\Vep}}$ (orange)  and  $\JDH{\UNIF}{{^8\LTV}}$ (blue) for inputs $\{0, 1/2, 1 \}$ {placed within the (triangular) probability simplex}.
      The blue points are within in the convex hull of the orange points.}     
    \end{subfigure}
  \caption{$N$-geometric and \TN-Laplace mechanisms.}
  \label{fig:lap_posteriors-b}
\end{figure}

The \TN-Laplace mechanisms approximate  $\LTV$ by $T$-pixelation of their outputs. Here $x$ is (still) in $\UI_N$ but $y$ is in  $\UI_{\kern.1emT}$. 
\begin{equation}\label{e1327a}
\begin{array}{lll}
^T\LTV(x)(y) \Wide{\Defs}
	&  \LTV(x)[y, y{+}\NF{1}{T})
		& ~\textit{if}~ y{<}1{-}\NF{1}{T} \\
	&  \LTV[1{-}\NF{1}{T}, 1]
		&~\textit{otherwise.}
\end{array}
\end{equation}
That is, we pixelate the \YY\ using $T$ for the Laplace (independently of the $N$
we use for \XX.) This is illustrated in \Fig{f1343a}.

Observe that as this is a post-processing (\Sec{s0931}(3)) of the output of $\LTV$, the refinement $\LTV\Ref ~^T\LTV$ follows.

\SSection{Refinement between $N$-Geometric\\ and \TN-Laplace mechanisms}
 
We now demonstrate the crucial fact that $\Geo_N^\Vep$ is refined by $^T\LTV$. We use version \Itm{i1327-b} of refinement, described in \Sec{s0931}, and establish a Wasserstein-style earth-move between hypers $\JDH{\UNIF}{{\Geo^\Vep_N}}$ and $\JDH{\UNIF}{{^T\LTV}}$ (i.e.\ for uniform prior \UNIF). 
 
\begin{lemma}\label{l1218}
For all integer $T{>}0$ we have that $\Geo_N^\Vep\Ref~^T\LTV$.

\begin{IEEEproof}
Take $\Delta, \Delta'$ in $\Dist^2 \UIK_N$ as hypers both with finite supports. We can depict such hypers in $\Real^{N+1}$-space by locating their supports, each of which is a point in $\Real^{N+1}$, where the axes of the diagram correspond to each point in $ \UI_N$. For example if we take $\Delta$ to be the hyper-distribution $\JDH{\UNIF}{{\Geo_2^\Vep}}$, it has three posterior distributions, which are 1-summing triples in $\Real^3$. They are depicted by the orange points in \Fig{fig:lap_posteriors-b}. Similarly the supports of the a hyper-distribution $\Delta'$ taken to be $\JDH{\UNIF}{{^T\LTV}}$  are represented by the blue dots.  Notice that the blue dots are contained in the convex hull of the orange dots, and this observation allows us to prove that the mechanisms $\Geo_2^\Vep$ and $^8\LTV$ are in a refinement relation.

We use the following fact \cite[Lem. 12.2]{Alvim:20a} about refinement $(\Ref)$.  

\begin{quotation}
Let $C, C'\In \UI_N\,{\MFun}\,\UIT$ be channels and let \UNIF\ be the uniform prior. 
  If the supports of $\JDH{\UNIF}{C}$ are linearly independent when considered as vectors in $\Real^N$, and their convex hull encloses the supports of  $\JDH{\UNIF}{C'}$, then $C\Ref C'$.\,%
\footnotemark
\end{quotation}
\footnotetext{The lemma applies to channels because of the direct correspondence between channels and the supports of hyper-distributions formed from uniform priors.}
To apply this result, we let $C$ be $\Geo_N^\Vep$ recall that indeed the supports of $\JDH{\UNIF}{{\Geo_N^\Vep}}$ are linearly independent (see for example \cite{Chatzi:2019}). Moreover in general, the supports of   $\JDH{\UNIF}{{^T\LTV}}$ are also contained in the convex hull. We provide details of this latter fact in \cite[Appendix {\textsection}B]{Optimality:FullPaper:2021}.
 \end{IEEEproof}
\end{lemma}

Finally we can show full refinement between the Laplace and the Geometric mechanism, which follows from continuity of refinement \cite{McIver:12}.

\begin{theorem}\label{t1555-a}
$\Geo_N^\Vep \Ref \LTV$.
\begin{IEEEproof}
We first form an anti-refinement chain $\dots{\Ref}~^{T_1} \LTV{\Ref}~^{T_0} \LTV$ so that (a) $\LTV {\Ref} ^{T_i} \LTV$ for all $i$, \emph{and} (b) the chain converges to $\LTV$.

We reason as follows:
\begin{Reason}
\Step{}
{\Geo_N^\Vep \Ref \LTV}
\WideStepR{iff}{$\Ref$ is continuous; (a), (b) above}
{\Geo_N^\Vep \Ref~^{T_i} \LTV\quad \textrm{~for all~} i{\geq}0} ~
\end{Reason}
which follows from \Lem{l1218}. We provide details of (a), (b) just above in \cite[Appendix {\textsection}B]{Optimality:FullPaper:2021}.
\end{IEEEproof}
\end{theorem}

We have shown that the Laplace mechanism is a refinement of the Geometric mechanism. This means that it genuinely leaks less information than does the Geometric mechanism and therefore affords greater privacy protections. On the other hand this means that we have lost utility with respect to the aggregated information. In the next section we turn to the comparison of the Laplace and Geometric mechanisms with respect to that loss.

\SSection{The Laplace approximates the Geometric}\label{ss957}
The geometrical interpretation of the Laplace and Geometric mechanisms set out above indicates how the Laplace approximates the Geometric as $\UIK_N$'s interval-width approaches $0$. In particular the refinement relationship  established in \Thm{t1555-a} describes how the posteriors
of $\JDH{\UNIF}{^{T} \LTV}$ all lie in between pairs of posteriors of  $\JDH{\UNIF}{\Geo^\Vep_N}$. This relationship between posteriors translates to a bound between the corresponding expected losses $ \Loss{\UNIF}{\LTV}{\ell}$ and $ \Loss{\UNIF}{\Geo_N^\Vep}{\ell}$ 
via the  Kantorovich-Rubinstein theorem applied to the hypers $\JDH{\UNIF}{^T\LTV}$ and $\JDH{\UNIF}{\Geo_N^\Vep}$. 
We sketch the argument in the next theorem, and provide full details in \cite[Appendix {\textsection}D]{Optimality:FullPaper:2021}.

\begin{theorem}\label{t1431}
Let $\ell$ be a $\kappa$-Lipschitz loss function, and $\UNIF$ the uniform distribution over $\UI_N$. Then
\begin{equation}\label{et1431}
 \Loss{\UNIF}{\LTV}{\ell}-\Loss{\UNIF}{ \Geo_N^\Vep}{\ell} \Wide{\leq} c\kappa/N~,
\end{equation}
where $c=3/(1{-}e^{-\Vep})^2$ is  constant for fixed $\Vep$. 
\begin{IEEEproof}
We appeal to the Kantorovich-{Rubinstein} theorem which states that the ``{Kantorovich} distance'' between probability distributions $\Delta, \Delta'$ bounds above the difference between expected values 
$|\Exp{\Delta}{f} -\Exp{\Delta'}{\kern-.2emf}| $
whenever $f$ satisfies the $\kappa$-Lipschitz condition.
 In our case the relevant distributions are the \emph{hyper}-distributions $\JDH{\UNIF}{^{T} \LTV}$ and $\JDH{\UNIF}{\Geo_N^\Vep}$, and the relevant Lipschitz functions are $Y_\ell$ for loss functions $\ell$.
 \footnote{
 Some $f\In \Dist\XX\Fun \Real$ is $\kappa$-Lipschitz if $|f(\delta)-f(\delta')|\leq \kappa \Kant(\delta, \delta')$, for $\kappa{>}0$, and $\Kant(\delta, \delta')$ is the Kantorovich distance between $\delta,\delta'$.}

We write $\Kant(\Delta, \Delta')$ for the Wasserstein distance between hyper-distributions $\Delta, \Delta'$ which is determined by the minimal \emph{earth-moving} cost to transform $\Delta$ to $\Delta'$. For any such earth move each posterior $\delta$ of $\Delta$
 is reassigned to a selection of posteriors of $\Delta'$ in proportion to the probability mass that $\Delta$  assigns to $\delta$. The cost of the move is the expected value of the distance between posterior reassignment (weighted by the proportion of the reassignment). Thus the cost of any specific earth move provides an upper bound to  $\Kant(\Delta, \Delta')$. 
\footnote{All the costs are determined by the underlying metric used to define the probability distributions. For us this is determined by the Euclidean distance on the interval $[0,1]$.}
Importantly for us, the relation of refinement $\Ref$ determines a specific earth move \cite{Alvim:20a} whose cost we can calculate.

Referring to \Lem{l1218} and \Fig{fig:lap_posteriors-b}, we see that the refinement between the approximation to the Laplace  $\JDH{\UNIF}{^T\LTV}$ and $\JDH{\UNIF}{\Geo_N^\Vep}$, reassigns the Geometric's posteriors (the orange dots) to the Laplace's posteriors (the blue dots). Crucially though the Geometric's posteriors form a linear order according to their distance from one another, and the refinement described in \Lem{l1218} shows how each Laplace posterior lies in between adjacent pairs of Geometric posteriors (according to the linear ordering), provided that $N$ divides $T$.
Therefore any redistribution of a Geometric posterior is bounded above by the distance to one or other of its adjacent posteriors. We show in \cite[Appendix {\textsection}D]{Optimality:FullPaper:2021} that distances between adjacent pairs is bounded above by $c/N$, and therefore $\Kant(\JDH{\UNIF}{^T\LTV}, \JDH{\UNIF}{\Geo_N^\Vep})\leq c/N$. 
 
Next we observe that if $\ell(w, x)$ is a $\kappa$-Lipschitz function on $[0, 1]$ (as a function of $x$), then $Y_\ell$ is a $\kappa$-Lipschitz function, and so by the Kantorovich-Rubinstein theorem we must have (recalling from \Eqn{e1536}) that $\Loss{\pi}{M}{\ell} {=}
\Exp{\JDH{\pi}{M}}{Y_\ell} $: 
\begin{equation}\label{e1638a}
 \Loss{\UNIF}{^T\LTV}{\ell}-\Loss{\UNIF}{\Geo_N^\Vep}{\ell} \Wide{\leq} c\kappa/N~.
\end{equation}
By \Thm{t1555-a} and post-processing we see that $\Geo_N^\Vep{\Ref}\LTV {\Ref} ~^T\LTV$. Recall from \Eqn{i1327-a} that refinement means that the corresponding losses are also ordered, i.e.\
\[
 \Loss{\UNIF}{\Geo_N^\Vep}{\ell} \leq  \Loss{\UNIF}{\LTV}{\ell} \leq  \Loss{\UNIF}{^T\LTV}{\ell}
\]
 and so the difference $\Loss{\UNIF}{\LTV}{\ell}-\Loss{\UNIF}{\Geo_N^\Vep}{\ell}$ must be no more than the difference $\Loss{\UNIF}{^T\LTV}{\ell}-\Loss{\UNIF}{\Geo_N^\Vep}{\ell}$ , thus  \Eqn{et1431} follows from \Eqn{e1638a}. Full details are set out in \cite[Appendix {\textsection}D]{Optimality:FullPaper:2021}.
\end{IEEEproof}
\end{theorem}

More generally, \Eqn{et1431} holds whatever the prior.

\begin{theorem}\label{t1431-b}
Let $\ell$ be a $\kappa$-Lipschitz loss function, and $\pi$ any prior over $\UI_N$. Then
\begin{equation}\label{et1431-b}
 \Loss{\pi}{\LTV}{\ell}-\Loss{\pi}{\Geo_N^\Vep}{\ell} \Wide{\leq} c\kappa/N~.
\end{equation}
\begin{IEEEproof}
This follows as for \Thm{t1431}, by direct calculation, noting that for discrete distributions we have   $ \Loss{\UNIF}{M}{\ell^*}=  \Loss{\pi_N}{M}{\ell}$, where $\ell^*(w, x) {\Defs}\ell(w, x){\times}\pi_N(x){\times N}$. Details are given in \cite[Appendix {\textsection}D]{Optimality:FullPaper:2021}.
\end{IEEEproof}
\end{theorem}

\SSection{Approximating monotonic functions}\label{ss956}
The final piece needed to complete our generalisation of the Ghosh et al.'s optimality theorem is monotonicity. We describe here how to approximate continuous monotonic functions, and expose the limitations for the Laplace mechanism.

\begin{definition}\label{d1158-b}
The loss function $\ell : {\cal W}{\times}{\cal X}\Fun \Real$ is said to be \emph{monotone} if:
there is some mapping $\theta\In {\cal W}{\Fun} [0,1]$,  such that
\[
\ell(w,x ) \Wide{\Defs} m(|\theta(w){-}x|,x)~,
\]
where $m: \Real{\times}\Real \Fun \Real$ is monotone in its first argument.
\end{definition}

Notice how $\theta$ takes care of any remapping that might need to be applied for computing expected losses. Interestingly step functions are not in general monotone on the whole of the continuous input $[0,1]$, but fortunately they are for the restrictions to $\UI_N$ that we need. 


\begin{lemma}\label{l1534-aa}
Let $\ell$  be  monotone. Then the approximation $\ell_T$ restricted to $\UI_N$ is monotone  whenever $T$ is a multiple of $N$.%
\begin{IEEEproof}
If $x {\in} \UIK_N$ then  $\NFloor{x}{T} {=} x$ since $N$ divides  $T$.
\end{IEEEproof}
\end{lemma}

Examples of continuous monotone loss functions include $\len$ and $\len^2$, where  $x, w \in [0,1]$, and 
\begin{equation}\label{e1538}
\len(w, x) \Wide{\Defs} |x{-}w|~.
\end{equation}
Note that $\len$ is 1-Lipschitz and $\len^2$ is 2-Lipschitz.

We note finally that as the pixellation of $N$ of $\ell$ increases the approximations $\ell_N$ converge to $\ell$.

\Section{Universal optimality \\ for the Laplace Mechanism}\label{s1352}
We finally have all the pieces in place to prove our main result,  \Thm{t1544} from \Sec{s160734} --- the generalisation of  discrete optimality \cite{Ghosh:12}.

Let $K^\Vep$ be any continuous \VDP\  mechanism with input $[0,1]$, and let $\pi$ be a (continuous) probability distribution over $[0,1]$ and $\ell$ a legal (i.e.\ continuous, monotone, $\kappa$-Lipschitz) loss function. Then: 
\begin{equation}\label{e1442}
\Loss{\pi}{\LTV}{\ell} \Wide{\leq} \Loss{\pi}{K^\Vep}{\ell}~.
\end{equation}
\begin{IEEEproof}
We use the above results to approximate the expected posterior loss by step functions; these approximations are equivalent to posterior losses over discrete mechanisms satisfying \VDP\ enabling appeal to Ghosh et al.'s universal optimality result on discrete mechanisms. 
We reason as follows: 


\bigskip


\begin{Reason}
\Step{}
{ \Loss{\pi}{K^\Vep}{\ell_N}\times e^{\nicefrac{\Vep}{N}}}
\StepR{$\geq$}{\Lem{l1720-a}}
{ \Loss{\pi_N}{K^\Vep_N}{\ell_N}}
\StepR{$\geq$}{\Lem{l1446}:  $\ell_N$ is legal by \Lem{l1534-aa}}
{ \Loss{\pi_N}{\Geo_N^\Vep}{\ell_N}}
\WideStepR{$\geq$}{\Thm{t1431-b}; $\ell_N$ is $\kappa$-Lipschitz}
{ \Loss{\pi_N}{\LTV}{\ell_N}-~c\kappa/N }
\StepR{$\geq$}{\Lem{l1720-a}}
{\Loss{\pi}{\LTV}{\ell_N}\times e^{\nicefrac{-\Vep}{n}}~-c\kappa/N~. }
\end{Reason}
\bigskip

The result now follows as above by taking $N$ to $\infty$, and noting that $e^{\nicefrac{\Vep}{N}}$, $e^{\nicefrac{-\Vep}{N}}$, $c\kappa/N$ and $\ell_N$ converge to $1,1,0,\ell$ respectively, and that taking expected values over fixed distributions is continuous.
\end{IEEEproof}

Note that \Thm{t1544} only holds for mechanisms that are \VDP. An arbitrary embedding $K_N$ is not necessarily \VDP, and in particular \Thm{t1544} does not apply to $\Geo_N^\Vep$.  Also the continuous property on $\ell$ is required, because $\ell_N$ must be monotone for all $N$. Thus arbitrary step functions do not satisfy this property, and so the Laplace mechanism is not in general universally optimal wrt.\ arbitrary step functions. Two popular loss functions however are continuous, and thus we have the following corollary.

\begin{corollary}\label{c1642}
The Laplace mechanism is universally optimal for $\len$ and $\len^2$.  
\end{corollary}

\Section{Related work}
The study of (universally) optimal mechanisms is one way to understand the cost of obfuscation, needed to implement privacy, but with a concomitant loss of utility of queries to databases.  Pai and Roth \cite{Pai:13} provide a detailed survey of the principles underlying the design of mechanisms including the need to trade utility with privacy, and Dinur et al.\ \cite{Dinur:03} explore the relationship between how much noise needs to be added to database queries relative to the usefulness of the data released. Our use of loss functions to measure utility follows both that of Ghosh et al.\ \cite{Ghosh:12} and Alvim et al.\ \cite{Alvim:2012aa}, and concerns optimality for entire mechanisms that satisfy a particular level of \VDP.  For example, the mean error $\len$ and the mean squared error $\len^2$ can be used to evaluate loss, as described by Ghosh et al.\ \cite{Ghosh:12} and mentioned in \Sec{ss956}.

The Laplace mechanism as a way to implement differential privacy has been shown for example by Dwork and Roth \cite{Dwork:2014c}. Moreover Chatzikokolakis et al.\ \cite{Chatzi2013} showed how it satisfied \VDP-privacy as formulated here using the Euclidean metric.

Whilst rare, optimality results avoid the need to design bespoke implementations of privacy mechanisms that are tailored to particular datasets. The Geometric mechanism appears to be special for discrete inputs, as Ghosh et al.\ \cite{Ghosh:12} showed when utility is measured using their ``legal'' loss functions. On the other hand, although the Laplace mechanism continues to be a popular obfuscation mechanism, its deficiencies in terms of utility have been demonstrated by others when the inputs to the obfuscation are discrete \cite{Soria-Comas:13}, and where the optimisation is based on minimising the probability of reporting an incorrect value, subject to the \VDP-constraint. Similarly Geng et al.\  \cite{Geng:15} show that adding noise according to a kind of  ``pixellated'' distribution  appears to produce the best utility for arbitrary discrete datasets. Such examples are consistent with our \Thm{t1555-a} showing where the Laplace mechanism is a refinement of the Geometric mechanism (loses more utility) when restricted to a discrete input (to the obfuscation). We mention also that optimal mechanisms have also been studied by Gupte et al.\ \cite{Gupte:10} wrt.\ minimax agents, rather than average-case minimising agents.

Other works have shown the Laplace mechanism is optimal for metric differential privacy in particular non-Bayesian scenarios. Koufogiannis et al.\ \cite{koufogiannis2015optimality} show that the Laplace mechanism is optimal for the mean-squared error function under Lipschitz privacy, equivalent to metric differential privacy; and Wang et al.\ \cite{wang2014entropy} show that the Laplace mechanism minimises loss measured by Shannon entropy, again for metric differential privacy. Our result on $\Real$ includes those results as specific cases; however, those works do go further in that they demonstrate optimality for their respective loss functions on $\Real^n$. We leave the study of these domains in the Bayesian setting to future work.

We also note that the linear ordering of the underlying query results seems to be important for finding optimality results.
For example  Brenner and Nissim \cite{Nissim:13} have demonstrated that for non-linearly ordered inputs, there are no optimal \VDP-mechanisms for a fixed level of $\Vep$. Although their result finds that only counting queries have optimal mechanisms, their context (oblivious mechanisms on database queries) does not include the possibility of continuous valued query results with a linear order; our result does not contradict their impossibility, it can be seen rather as an extension of this result to a continuous setting.

Alvim et al.\ \cite{DBLP:journals/jcs/AlvimACDP15} also use the framework of Quantitative Information Flow to study the relationship between the privacy and the utility of \VDP\ mechanisms. In their work they model utility in terms of a leakage measure, where leakage is defined as the ratio of input gain to output gain wrt.\ a mechanism modelled as a channel. Their \emph{gain} is entirely dual to our \emph{loss} here, and is a model of an adversary trying to infer as much as possible about the secret input. Other notions of optimality have also been studied in respect of showing that the Laplace mechanism is not optimal, including \cite{Asi:20} who work with ``near instance'' optimality, and Geng and Viswanath \cite{Geng:15} show how to scale the Laplace in various ways to obtain good utility. Note also that these definitions of optimality do not use a prior, and therefore represent the special case of utility per exact input, rather than in a scenario where the observer's prior knowledge is included.

The use of the Laplace mechanism in real privacy applications has been demonstrated by Chatzikokolakis et al.\ \cite{Chatzi2013}
for geolocation privacy, and \cite{Wang:13} for  for privacy-preserving graph analysis, and Phan et al.\ \cite{Phan:17}  
in deep learning. 

Information-theoretic channel models for studying differential privacy were originally proposed by Alvim et al.\ \cite{DBLP:conf/icalp/AlvimACP11,DBLP:journals/jcs/AlvimACDP15}, and extended to arbitrary metrics in \cite{Chatzi2013}.

\Section{Conclusion}

We have studied the relationship between differential privacy (good) and loss of utility (bad)  when the input \XX\ can be over an interval of the reals, rather than having \XX\ described as in the optimality result of Ghosh et al.\ \cite{Ghosh:12,Ghosh:09}, i.e.\ as a discrete space.
Here we have instead used as input space the continuous interval $[0,1]$; but we note that the result extends straightforwardly to any finite interval $[a, b]$ of $\Real$. Our result also imposes the condition that the loss functions must be $\kappa$-Lipschitz for some finite $\kappa$. We do not know whether this condition can be removed in general. 

We observe that for \NStep\ loss functions, the Laplace mechanism is not optimal, and in fact a bespoke Geometric mechanism \emph{will} be optimal for such loss functions. However our \Thm{t1431-b} provides a way to estimate the error, relative to the optimal loss, when using the Laplace mechanism.

Finally we note that the space of \VDP\ mechanisms is very rich, even for discrete inputs, suggesting that the optimality result given here will be useful whenever the input domain can be linearly ordered.

\subsection*{Acknowledgements}

We thank Catuscia Palamidessi for suggesting this problem to us.

\appendix
The appendices may be found at \cite{Optimality:FullPaper:2021}.

\newpage
\bibliographystyle{IEEEtran}

\end{document}